\documentclass{ckm}                 % twocolumn proceedings style
\confname{Workshop on the CKM Unitarity Triangle, IPPP Durham, April
  2003}

\title{Performance of CDF for $B$ physics}

\author{
R.G.C. Oldeman\\
for the CDF collaboration}
\address{University of Pennsylvania, Philadelphia}

\begin{document}

\begin{abstract}
Hadron colliders can be an abundant source of heavy flavor quarks, 
but pose a challenge to isolate the physics signals from the high backgrounds.
The upgraded CDF\,II detector,
with its precise tracking capabilities and powerful trigger system,
is well equipped for this task.
The detector is described with an emphasis on actual performance
and on techniques to maximize the heavy flavor yield.
Some first heavy flavor results are summarized.
\end{abstract}
\maketitle

\section{Introduction}

In Run\,I of the Tevatron, CDF made essential contributions to $B$ physics,
providing some of the best measurements of masses, lifetimes, mixing and branching ratios.
The $B$ factories are now setting new standards with a wealth of precision measurements,
posing stringent tests of the standard model.
CDF has made significant improvements of the detector and the trigger
to expand its capabilities beyond simply profiting from 
the higher luminosities of the Tevatron in Run\,II. 

Both $B$ and $D$ mesons are amply produced at the Tevatron:
in Run\,I the $B^+$ cross section with transverse momentum \mbox{$p_T\geq6$\,GeV/$c$}
and rapidity \mbox{$|y|\leq1$}
was measured in the \mbox{$B^+\to J/\psi K^+$} mode to be \mbox{$3.6\pm0.6\,\mu$b}~\cite{Acosta:2001rz}.
A new preliminary Run\,II measurement of the charm meson cross sections
for the same kinematic region yields
even larger values: \mbox{$4.3\pm0.7\,\mu$b}
for $D^+$ and \mbox{$9.3\pm1.1\,\mu$b} for $D^0$, 
both measured using fully reconstructed hadronic decays.

Heavy flavor cross sections at hadron colliders are three orders of magnitude larger than 
at $e^+e^-$ machines running at the $\Upsilon(4S)$,
and include the heavier $B_s$ and $B_c$ mesons
and the $\Lambda_b$ and $\Xi_b$ baryons.
The challenge at a hadron collider is to select in real time 
the heavy flavor events from the overwhelming QCD background.
Therefore the emphasis in this talk is on the trigger system,
which is a particular strength of CDF\,II.
% with respect to other experiments at hadron colliders.

\section{Performance of the Tevatron}
%The Tevatron is a 6\,km circumference 4.5\,T superconducting proton-antiproton collider.
The center-of-mass energy of the Tevatron has increased 
from 1.8\,TeV in Run\,I to 1.96\,TeV in Run\,II, and
the number of proton and anti-proton bunches has increased from 6 to 36,
with a bunch separation of 396\,ns.
We anticipate the number of interactions per bunch crossing  to rise from 1.2
at present peak luminosities of \mbox{$4\times10^{31}$\,cm$^{-2}$s$^{-1}$}
to 6 at future luminosities of \mbox{$2\times10^{32}$\,cm$^{-2}$s$^{-1}$}.
This number can be reduced by increasing the number of bunches, but
the upgrade to 132\,ns bunch spacing has been indefinitely postponed.
At the time of this workshop, the Tevatron has delivered 190\,pb$^{-1}$,
of which CDF has recorded 145\,pb$^{-1}$ on tape. 
Present $B$ physics analyses use up to 70\,pb$^{-1}$ of this data,
where part of the data was not used because the silicon detector was not
fully operational.

\section{The CDF\,II detector}

CDF\,II~\cite{Blair:1996kx} has the typical structure of a 4$\pi$ detector at a symmetric collision point.
A superconducting solenoid provides a uniform 1.4\,T field over a tracking volume of 18\,m$^3$.
Closest to the beam is a silicon system to 
provide an accurate measurement of the impact parameter of charged tracks.
A large-radius, low-material drift chamber gives accurate
measurements of tens of charged particles per interaction.
New in Run\,II is a Time-of-Flight (TOF) detector for particle identification at low momenta.
Outside the magnet are the calorimeters followed by muon identification systems.
The components most relevant for $B$ physics are described in more detail below.

\subsection{The Silicon system}
CDF\,II has three silicon systems~\cite{Sill:zz}. 
Layer 00 (L00) is a single-sided low-mass radiation-hard silicon detector,
mounted on the beryllium beam pipe at a radius of 1.6\,cm. 
The SVX\,II is a five-layer double-sided silicon strip detector.
It consists of 3 barrels, each 30\,cm in length and covers 90\% of the extended interaction region
(RMS $\approx$\,30\,cm).
The Intermediate Silicon Layer (ISL) provides one additional 
double-sided layer in the central region and two in the forward and backward regions,
to improve track reconstruction.
A light-weight carbon-fiber space frame keeps the 3.5\,m$^2$ 
of ISL sensors stably aligned at the 10\,$\mu$m level.
All silicon sensors use the same 128 channel SVX\,III readout chip
that uses an analog pipeline for deadtimeless operation. 
The silicon system has $\approx$750k channels in total and can be readout in 20\,$\mu$s.

\subsection{Commissioning the silicon detector}
In the first year of data-taking for Run\,II, CDF has faced many startup problems
with the silicon system that have now mostly been overcome.
The SVX\,II has suffered from several beam incidents where 
a relatively small amount of radiation was incident on the silicon sensors 
and chips within a few nanoseconds, 
resulting in the loss of several sensors.
The Tevatron has now installed interlocks to prevent this class of beam incidents.
Other damage occurred when the silicon detector was read out at fixed 
frequencies of several kHz during anomalous trigger conditions.
This has been traced down to resonant oscillations of wirebonds
perpendicular to the 1.4\,T magnetic field.
CDF has now implemented hardware protection against fixed-frequency readout.
The ISL suffered from insufficient cooling, as 11 of the 12 cooling lines were
internally blocked with epoxy at some of the 90$^\circ$ elbow connections.
Using boroscopes to find the blockages and a pulsed 40\,W laser to destroy them,
10 of the 11 blocked lines have now been cleared.
L00 is affected by  pickup from the digital readout lines on the analog signal cables.
Methods have been developed to correct for this in software, 
but they require non-sparsified readout, prohibiting the use of L00 in the trigger.
The CAEN power supplies that supply the L00 sensors with the bias voltage
suffered from beam related failures.
This has been solved by replacing the MOSFET voltage regulator by a BJT circuit.
At the time of this workshop more than 90\% of the silicon detector is operational
and used in physics analyses.
One disadvantage of the present silicon detector is the large amount of passive
material inside the tracking region:
a track traverses on average $\approx$15\% of a radiation length.

\subsection{The Central Outer Tracker (COT)}
The Central Outer Tracker~\cite{Pitts:qy} is a cylindrical Ar/Et drift chamber based
on the highly successful CTC used in Run\,I, 
modified to cope with the shorter time between interactions.
The COT has 96 layers,
organized in 8 super layers, alternating between axial and $\pm2^\circ$ stereo readout.
The single-wire resolution is approximately 200\,$\mu$m, resulting in a 
momentum resolution of \mbox{$\frac{\Delta p_T}{p_T}\approx(0.7\oplus0.1p_T)\%$}.
The measurement of the specific energy deposition $dE/dx$ gives a 
$\pi/K$ separation of $\approx$1.2\,$\sigma$ in the region of the relativistic rise ($p_T\geq2$\,GeV).
The COT has proven to be a very reliable detector, with fewer than 1\% dead channels.
The tracking efficiency for high-$p_T$ isolated tracks has been measured
from data using \mbox{$W^+\to e^+ \nu_e$} events to be better than 99.5\%.

\subsection{The Time-of-Flight detector}
For the first time at a hadron collider experiment,
CDF\,II has installed a Time of Flight (TOF) detector for particle identification.
It consists of 216 scintillator bars, 3\,m long and $4\times4$\,cm 
in cross section, mounted between the COT and the superconducting solenoid.
The bars are read out on both sides by compact fine-mesh 19-stage photomultipliers,
that allow operation inside the full 1.4\,T magnetic field.
The achieved timing resolution is 120\,ps, close to the 100\,ps design,
giving 2$\sigma$ separation between pions and kaons up to momenta of 1.5\,GeV/$c$.
The TOF has been designed primarily to improve $B$ flavor tagging,
in particular for the $B_s$ meson, which is often produced
together with a kaon, correlated in charge to the $b$ quark produced.
The TOF has also found use for proton identification, 
cosmic ray rejection and monopole searches.
The TOF performs reliably and all 432 channels are fully functional.
However, the occupancy of approximately 20\% is about two times higher than expected, 
which has affected the reconstruction efficiency.

\section{The three-level trigger}
CDF uses a three-level trigger system to reduce the 1.7\,MHz bunch crossing rate
to 75\,Hz written on tape.
The first trigger level is deadtimeless, with a 42 buffer deep pipelined system,
allowing 5.5\,$\mu$s to form a trigger decision.
The eXtremely Fast Tracker (XFT)~\cite{Thomson:2002xp} 
reconstructs COT tracks in the axial projection down to transverse momenta of 1.5\,GeV/$c$,
using hits with coarse timing information.
Muon and electron signals can also be matched to tracks. 
Information about energy deposition in the hadron calorimeter 
are available at Level\,1, but are not used for $B$ triggers.
The maximum input rate for Level\,2 is presently about 20\,kHz.
We expect this to reach 25\,kHz soon, but the original design rate of 50\,kHz 
will be difficult to reach.
At Level\,2, the Silicon Vertex Tracker (SVT)~\cite{Ashmanskas:1999ze} provides a list of tracks
using axial hits of 4 silicon layers.
This allows to select displaced tracks, which greatly enhances the purity of heavy flavor
signals. 
Typical rate reductions are O(10$^{-1}$) when requiring one track with a 
displacement of $\geq$100\,$\mu$m, or  O(10$^{-2}$) when requiring two displaced tracks.
Further trigger rate reductions are achieved when selecting on the opening angle between two tracks, 
requiring positive lifetime or pointing of the two-track momentum vector to the primary vertex.
The readout time of the COT currently limits
the rate into the third trigger level to less than 300\,Hz.
At Level\,3, the complete events are processed on a farm of
300 PCs, which perform full event reconstruction
and write up to 20\,MB/s to tape.
The event size of almost 300\,kB limits the Level\,3 output
rate to less than 75\,Hz, but data compression techniques are being developed to increase this
to more than 100\,Hz. 

\section{Trigger strategies}
The vast majority of $B$ physics at CDF is based on three sets of triggers.
The \emph{dimuon} trigger requires two muons with \mbox{$p_T\geq1.5$\,GeV/$c$}.
It collects vast samples of \mbox{$J/\psi\to\mu^+\mu^-$} decays, which 
come in about 15\% of the cases from a $B$ decay. 
At present peak luminosities this trigger uses 40\,Hz at Level\,2
and new hardware is being built to
select on the transverse mass of a muon pair, which should reduce this rate significantly.
The \emph{lepton+displaced track} trigger uses muons and electrons with \mbox{$p_T\geq4$\,GeV/$c$}
combined with a displaced track found by the SVT.
The specific yield of semileptonic $B$ decays with this trigger is 5 times
higher than in the Run\,I equivalent, which was based on an 8\,GeV/$c$ lepton.
These signals are used for lifetime measurements
and for optimizing flavor tagging algorithms.
The \emph{two-track} trigger  requires two displaced tracks.
One version requires a large opening angle  
and the reconstructed $B$ should point back to the primary vertex.
It is designed to collect 2-body $B$ decays, such as \mbox{$B_d\to\pi^+\pi^-$}. 
The other requires smaller opening angles and has stricter requirements on the track displacement
(120\,$\mu$m instead of 100\,$\mu$m). 
This second version is designed for multi-body $B$ decays, such as \mbox{$B_s\to D_s^- \pi^+$};
it has also proven useful for collecting huge samples of
hadronic $D$ meson decays.
The two-track trigger uses almost all of the available Level\,1 bandwidth 
and a significant fraction of the bandwidth at Level\,3.
We are constantly trying to reduce the trigger rates
while keeping the maximum amount of real heavy flavor events.

\section{Techniques for optimizing heavy flavor yields}
\subsection{Dynamic prescaling}
\begin{figure}
\hbox to\hsize{\hss
\includegraphics[width=\hsize]{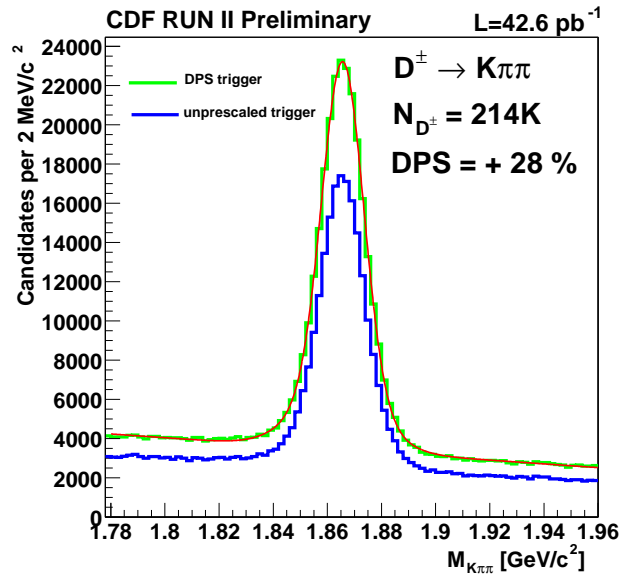}
\hss}
\caption{Event yield of \mbox{$D^+\to K^-\pi^+\pi^+$} since the introduction of dynamic
prescaling. 
Blue curve is the $D^+$ yield of the nominal trigger,  
the green curve is the $D^+$ total yield including the low-$p_T$ trigger
that was effectively prescaled by a factor 0.22.}
\label{fig:dps}
\end{figure}

A typical ``store'' at the Tevatron lasts about 20 hours, 
during which time the luminosity has decreased by a factor 2.5 with respect to the
initial luminosity. 
Since the trigger bandwidth has been designed for the peak luminosity, 
significant bandwidth is available in the later part of a store.
We have developed techniques to fill this available bandwidth with $B$ physics
by implementing triggers with several different $p_T$ thresholds and 
increasing the acceptance fraction (prescale) of the triggers with lower thresholds
as the luminosity decreases.
The two-track trigger comes in three versions. The \emph{high-$p_T$} trigger
has a low rate due to high thresholds  
(two oppositely charged tracks of \mbox{$p_T\geq2.5\,$GeV/$c$} with \mbox{$\Sigma p_T\geq6.5$\,GeV/$c$})
and is never prescaled.
The rate is low enough that it can be safely run up to luminosities of \mbox{$8\times10^{31}$\,cm$^{-2}$s$^{-1}$}.
The \emph{nominal} trigger requires two oppositely charged tracks
of \mbox{$p_T\geq2.0\,$GeV/$c$} with \mbox{$\Sigma p_T\geq5.5$\,GeV/$c$}.
This trigger is prescaled when the Level\,1 rate exceeds 20\,kHz.
A \emph{low-$p_T$} trigger that requires two tracks
of \mbox{$p_T\geq2.0\,$GeV/$c$} and no $\Sigma p_T$ or opposite charge requirement
is enabled when the Level\,1 rate drops below 16\,kHz.
In the six months that dynamic prescaling has been enabled, 
the low-$p_T$ trigger has given a yield increase of 28\% in the \mbox{$D^+\to K^-\pi^+\pi^+$} channel,
as shown in Figure~\ref{fig:dps}, while it was prescaled on average by a factor 4.5.

\subsection{XFT 1-miss configuration}
Until October 2002, the XFT required 10 out of 12 hits on the four axial superlayers of the COT
to reconstruct a track. 
This resulted in an efficiency of $>$95\%, 
but with a purity that was considered insufficient.
Studies showed that the COT single hit efficiency was high enough
that requiring 11 out of 12 hits resulted in an efficiency still better than 90\%,
while reducing the rate of the two-track trigger by almost a factor two
at high luminosity.

\subsection{4/5 SVT tracking}
\begin{figure}
\hbox to\hsize{\hss
\includegraphics[width=\hsize]{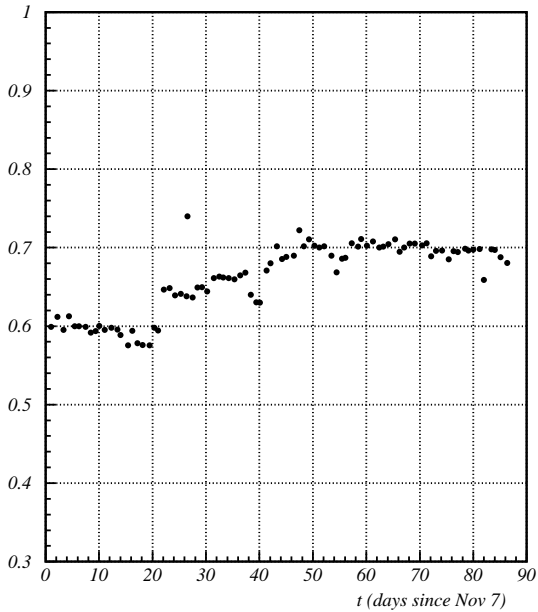}
\hss}
\caption{Efficiency of the SVT in days since Nov 7, 2002.
The rise in efficiency from 60\% to 70\% comes from improved SVX coverage.
The high point at $t=26$ corresponds to a test run with 4/5 SVT tracking.}
\label{fig:effsvt_time}
\end{figure}

The Silicon Vertex Tracker requires the presence of four strip clusters
in four layers, resulting in an overall efficiency that scales as the
fourth power of the single hit efficiency.
It is possible to program the SVT to use all five SVX layers,
and allow a missing hit in one layer.
In a test-run, we found a 15\% increase of the SVT single-track efficiency,
as shown in Figure~\ref{fig:effsvt_time}.
Additional gain comes from the efficiency improvement for tracks
that cross between SVX half-barrels.  
We find an increase of the heavy flavor yield in the 
two-track trigger of more than 50\%.
However, the trigger rate at Level\,2 rises by almost 100\% 
and we are studying how to improve the purity.

\subsection{Silicon tracking at Level\,3}
While the core of the impact parameter resolution of the SVT is similar 
to the offline resolution (both $\approx$50\,$\mu$m, including a contribution of 
$\approx$30\,$\mu$m from the beamspot),
the SVT impact parameter measurement has significant non-Gaussian tails, 
which can be attributed to the handling of large clusters of overlapping hits in the
silicon detector.
The reconstruction software at Level\,3 
can reduce the non-Gaussian resolution tails
because more time and more information 
are available for complex pattern recognition
and the handling of overlapping hits.
In May 2003, silicon tracking at Level\,3 
was implemented and resulted in a factor 2 reduction of
the rate written to tape of the two-track trigger with a small ($\approx$5\%) loss of 
heavy flavor signal.

\section{Some recent results}
The low $p_T$ threshold of the dimuon trigger has made it possible to collect 
$J/\psi$ decays down to zero transverse momentum.
We measured a cross section 
\mbox{$\sigma(J/\psi)_{|y|\leq0.6}$} $\times$ \mbox{$B(J/\psi \to \mu\mu)$} = 
\mbox{$240\pm1(stat)^{+35}_{-28}(syst)$\,nb}.
Using fully reconstructed \mbox{$B\to J/\psi X$} decays
%$B^+\to J/\psi K^+$, $B^0\to J/\psi K^{*+}$ and $B_s\to J/\psi\phi$,
we measure:
\mbox{$\tau(B^+)=1.57\pm0.07(stat)\pm0.02(syst)$\,ps},
\mbox{$\tau(B^0)=1.42\pm0.09(stat)\pm0.02(syst)$\,ps}, and
\mbox{$\tau(B_s)=1.25\pm0.20(stat)\pm0.02(syst)$\,ps}.

In the two-track trigger, we have reconstructed 450k \mbox{$D^0\to K^-\pi^+$} decays.
We looked for $D^0\to\mu^+\mu^-$ decays, found zero events on a background of 1.7,
and set a preliminary limit of \mbox{$2.7\times10^{-6}$} at 90\% CL.
We also looked for direct CP violation in $D^*$-tagged \mbox{$D^0\to\pi^+\pi^-$}
and \mbox{$D^0\to K^+K^-$} decays and find 
\mbox{$A_{CP,KK}=2.0\pm1.7(stat)\pm0.6(syst)$\%} and
\mbox{$A_{CP,\pi\pi}=3.0\pm1.9(stat)\pm0.6(syst)$\%}.

%In 65\,pb$^{-1}$ we find 300 $B\to h^+h^-$ decays, 
%and 44 $B_s\to D_s^- \pi^+$ decays, where $D_s^-\to\phi$, $\phi\to K^+K^-$.

\end{document}